\documentclass[%
 reprint,
 amsmath,amssymb,
 aps,prl
]{revtex4-2}

\usepackage{graphicx}       % Include figure files
\usepackage{dcolumn}        % Align table columns on decimal point
\usepackage{bm}             % Bold math
\usepackage{siunitx}        % SI units
\DeclareSIUnit{\gauss}{G}   % Define custom SI unit
\usepackage{physics}        % Physics package
\usepackage[english]{babel} % Language settings

\usepackage{xparse}         % For defining commands
\usepackage{color}          % Color package
\usepackage{xcolor}         % Extended color support
\usepackage{comment}        % Commenting out large sections
\usepackage{import}         % Importing files from different directories
\usepackage{hyperref}       % Hyperlinks
\usepackage{soul}           % For strikeout in editing

\NewDocumentCommand{\NSTIRAP}{}{\ensuremath{N_{\text{STIRAP}}}}

\NewDocumentCommand{\Tadi}{}{\ensuremath{\tau_{\text{adi}}}}
\NewDocumentCommand{\Tdeph}{}{\ensuremath{\tau_{\text{deph}}}}

\begin{document}

\preprint{APS/123-QED}

\title{Enhanced quantum state transfer via feedforward cancellation of  optical phase noise}

% Force line breaks with \\
%

\author{Benjamin P. Maddox}
\thanks{These authors contributed equally to this work.}
\author{Jonathan M. Mortlock}
\thanks{These authors contributed equally to this work.}
\author{Tom R. Hepworth}
\author{\mbox{Adarsh P. Raghuram}}
\author{Philip D. Gregory}
\author{Alexander Guttridge}
\author{Simon L. Cornish}
\email{s.l.cornish@durham.ac.uk}
\affiliation{%
Department of Physics, Durham University, South Road, Durham DH1 3LE, United Kingdom}%

\date{\today}% It is always \today, today,
\begin{abstract}

Many experimental platforms for quantum science depend on state control via laser fields. Frequently, however, the control fidelity is limited by optical phase noise. This is exacerbated in stabilized laser systems where high-frequency phase noise is an unavoidable consequence of feedback. Here we implement an optical feedforward technique to suppress laser phase noise in the STIRAP state transfer of ultracold RbCs molecules, across \SI{114}{\tera \hertz}, from a weakly bound Feshbach state to the rovibrational ground state. By performing over 100 state transfers on single molecules, we measure a significantly enhanced transfer efficiency of 98.7(1)\% limited only by available laser intensity.

\end{abstract}
\maketitle

Robust control of quantum states with optical fields is vital in many areas of modern physics. Depending on the platform, this can be achieved with one-photon or two-photon driving fields to perform both single-qubit and entangling operations \cite{gaeblerHighFidelityUniversalGate2016,levineHighFidelityControlEntanglement2018,schineLonglivedBellStates2022}. Control fidelities can be enhanced by using pulse shaping schemes \cite{torosovCoherentControlTechniques2021}. One widely used technique is Stimulated Raman adiabatic passage (STIRAP) \cite{bergmannCoherentPopulationTransfer1998,vitanovStimulatedRamanAdiabatic2017}, which enables the transfer of population between two discrete states via coupling to an intermediate state. Notable advantages of STIRAP are that it is immune to loss through spontaneous emission from the intermediate state, and it is relatively insensitive to noise in experimental conditions such as laser intensity~\cite{vitanovStimulatedRamanAdiabatic2017}. This has led to STIRAP finding important applications in superconducting circuits~\cite{kumarStimulatedRamanAdiabatic2016}, trapped ions~\cite{mollerEfficientQubitDetection2007}, nitrogen-vacancy centres~\cite{golterOpticallyDrivenRabi2014}, optomechanical resonators~\cite{fedoseevStimulatedRamanAdiabatic2021}, optical waveguides~\cite{longhiCoherentTunnelingAdiabatic2007} and ultracold molecule synthesis~\cite{koch_coherent_2012}. 

\begin{figure}[ht]
    \centering
    \includegraphics[width=\linewidth]{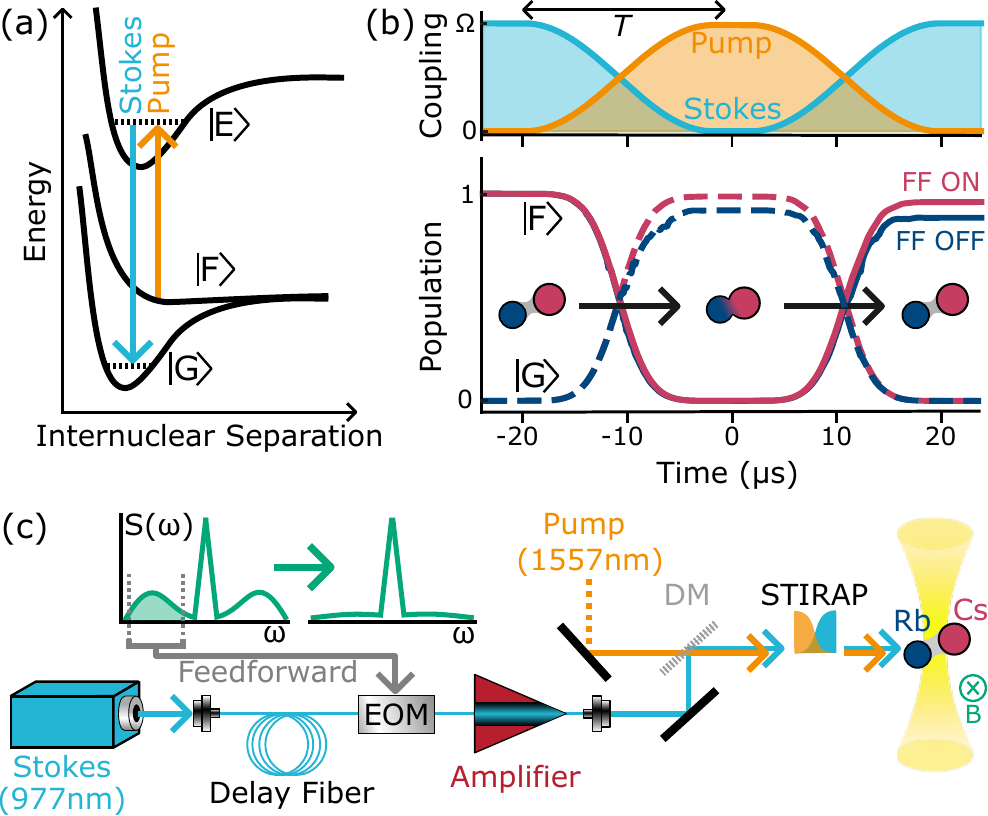}
    \caption{STIRAP with feedforward (FF) in ultracold RbCs molecules. 
    (a)~Schematic of the STIRAP states and transitions. Molecules are initially prepared and later detected in $\ket{F}$. 
    (b)~Pulse scheme for two-way STIRAP ($N_\mathrm{STIRAP}=2$) and the associated simulated populations of \(\ket{F}\) (solid lines) and \(\ket{G}\) (dashed lines). The red (blue) lines indicate the expected transfer with (without) the reduction in laser phase noise from FF.
    (c)~A simplified overview of the experiment. STIRAP of RbCs molecules in a tweezer array is achieved using two lasers, where FF noise cancellation is applied to each laser independently using a fiber EOM. The light is then amplified and combined before being sent to the molecules. The direction of the bias magnetic field (B) is orthogonal to the propagation direction of the STIRAP beams as shown. }
    \label{fig:concept}
\end{figure}

Despite being less sensitive to laser amplitude noise, STIRAP is inherently sensitive to fast laser phase noise as it relies on the adiabatic evolution of a dark state~\cite{vitanovStimulatedRamanAdiabatic2017,yatsenkoDetrimentalConsequencesSmall2014}. 
To minimise phase noise, lasers with narrow linewidths are required. This is commonly achieved by actively stabilising the frequency of the light to a stable reference such as an optical cavity. This process reduces phase noise at frequencies within the bandwidth of the feedback loop, but can also introduce additional noise at higher frequencies. This high-frequency phase noise is colloquially known as the \emph{servo bump}. As is generally the case in optical quantum control \cite{jiangSensitivityQuantumGate2023}, STIRAP is most affected by phase noise at frequencies comparable to the Rabi frequencies of the driving fields \cite{yatsenkoDetrimentalConsequencesSmall2014}. In many experiments, the servo bump and the driving Rabi frequency are unavoidably close together making this a challenge for high-fidelity control.

Filtering out the servo bumps may be achieved by passing the light through one or more additional optical cavities~\cite{haldEfficientSuppressionDiodelaser2005a,nazarovaLowfrequencynoiseDiodeLaser2008,akermanUniversalGatesetTrappedion2015,levineHighFidelityControlEntanglement2018,bauseEfficientConversionClosedchanneldominated2021}. This technique is effective, but the filtering is accompanied by a large loss in optical power. 
A recently reported technique based on feedforward noise cancellation has demonstrated a reduction in optical phase noise without cavity filtering \cite{liActiveCancellationServoInduced2022,chao_pounddreverhall_2024}. This approach measures the variation of phase in real time and then corrects the light sent to the experiment using an electro-optic modulator (EOM). This method is significantly simpler to implement and bypasses the power limitations associated with filter cavities.

Here we deploy feedforward phase-noise cancellation to significantly improve quantum state transfer, using STIRAP in ultracold molecules as a testbed for the technique. State transfer is an integral part of both the formation and detection of ground-state molecules produced by associating atoms. Efficient STIRAP is critical in applications where detection of the quantum state of individual molecules is required, for example in quantum simulation of spin models 
\cite{cornishQuantumComputationQuantum2024a,
gorshkovTunableSuperfluidityQuantum2011,
christakisProbingSiteresolvedCorrelations2023,
liTunableItinerantSpin2023,
bilitewskiDynamicalGenerationSpin2021,
yanObservationDipolarSpinexchange2013,
picardSubmillisecondEntanglementISWAP2024} 
or quantum information storage 
\cite{parkSecondscaleNuclearSpin2017,gregoryRobustStorageQubits2021}. 
Improved state transfer using STIRAP would also be highly beneficial in providing increased molecule number in other experiments, for example studying strongly dipolar degenerate gases 
\cite{schindewolfEvaporationMicrowaveshieldedPolar2022,
bigagliObservationBoseEinstein2024,
demarcoDegenerateFermiGas2019,
schmidtSelfboundDipolarDroplets2022} 
or precision measurement of fundamental constants 
\cite{safronovaSearchNewPhysics2018,demilleQuantumSensingMetrology2024,pandaStimulatedRamanAdiabatic2016,yeEssayQuantumSensing2024a}. 
% Recent progress in the control of collisions has enabled quantum degenerate gases \cite{demarcoDegenerateFermiGas2019,schindewolfEvaporationMicrowaveshieldedPolar2022,bigagliObservationBoseEinstein2024} needed for ground state quantum simulations.
% Quantum degenerate gases of fermionic~\cite{demarcoDegenerateFermiGas2019,schindewolfEvaporationMicrowaveshieldedPolar2022} and bosonic~\cite{bigagliObservationBoseEinstein2024} polar molecules have been produced in this way.
% Schemes for quantum computing and quantum simulation have been proposed for such molecules~\cite{cornishQuantumComputationQuantum2024}, but current STIRAP efficiencies will prevent experiments from achieving competitive error rates. Improving the efficiency of STIRAP is also important more broadly to the field of ultracold molecules. For example, it is used in state preparation in searches for the electron electric dipole moment \cite{pandaStimulatedRamanAdiabatic2016}.  

In this letter, we demonstrate the use of feedforward (FF) to significantly improve quantum state transfer using STIRAP. Specifically, we consider the transfer of ultracold $^{87}$Rb$^{133}$Cs (RbCs) molecules between a weakly bound Feshbach state and the absolute ground state spanning an energy gap of $\sim\!h \times\SI{114}{\tera \hertz}$ \cite{molonyMeasurementBindingEnergy2016}. We measure a transfer efficiency of 98.7(1)\% by performing over 100 one-way transfers on a single molecule in an optical tweezer; to our knowledge this is the highest reported transfer efficiency in any ultracold polar molecule. We are able to reproduce our results by simulating the state transfer Hamiltonian with realistic laser phase noise and magnetic field instability based on independent measurements in the experiment. The model shows that our current efficiency is only limited by the available laser power, and efficiencies approaching 99.9\% are possible with realistic laser intensities.

\begin{figure}[t!]
    \centering
    \includegraphics[width = \linewidth]{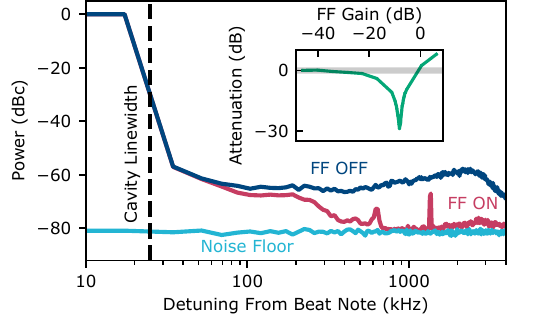}
    \caption{Self-heterodyne measurement of the Stokes laser with the cavity transmission. Noise power relative to the beat note carrier is shown for feedforward off (dark blue) and feedforward on (red). We also plot the detection noise floor for reference (light blue), measured by blocking the cavity transmission.
    % Noise at frequencies above the cavity linewidth can be considered exclusive to the main laser path. 
     Residual spikes in the FF on spectrum are believed to originate from interference due to neighbouring instruments. Inset: measured suppression of sinusoidal frequency modulation as a function of the gain of the FF amplifier, as described in the main text.}
    \label{fig:beatnote}
\end{figure}

Using STIRAP as the state transfer process for preparation and readout of ultracold molecules is generic to all species formed from ultracold atoms \cite{
niHighPhaseSpaceDensityGas2008,
takekoshiUltracoldDenseSamples2014,
molonyCreationUltracoldRb2014,
parkUltracoldDipolarGas2015,
guoCreationUltracoldGas2016,
rvachov_long-lived_2017,
seesselbergModelingAdiabaticCreation2018,
yang_observation_2019,
hu_direct_2019,
vogesUltracoldGasBosonic2020,
cairncrossAssemblyRovibrationalGround2021,
rosenberg_observation_2022,
stevensonUltracoldGasDipolar2023,
he_efficient_2024}.
% \jonathan{Perhaps there is a better ref}. 
Molecules are prepared in a weakly bound \emph{Feshbach} state $\ket{F}$, which is coupled in a \(\Lambda\) scheme to the target ground state $\ket{G}$, via a short-lived electronically excited state $\ket{E}$. We name the laser coupling to $\ket{F}$ \emph{pump} and that coupling to $\ket{G}$ \emph{Stokes}, as illustrated in Fig.\,\ref{fig:concept}(a). With the molecules prepared in $\ket{F}$, pulsing on the off-resonant Stokes laser first initialises the system in the dark state $\ket{D}=\cos{\theta}\ket{F} - \sin{\theta}\ket{G}$. Here, $\theta=\arctan{(\Omega_\mathrm{P}/\Omega_\mathrm{S})}$ is the mixing angle which depends on the ratio of the pump Rabi frequency $\Omega_\mathrm{P}$ and Stokes Rabi frequency $\Omega_\mathrm{S}$. 
Smoothly ramping down $\Omega_\mathrm{S}$ while ramping up $\Omega_\mathrm{P}$ preserves this dark state; this coherently transfers the molecules from $\ket{F}\rightarrow\ket{G}$, as shown in Fig.\,\ref{fig:concept}(b), without populating $\ket{E}$.

Lossless transfer requires extinguishing the matrix element that couples the dark state to the bright state~\cite{yatsenkoStimulatedRamanAdiabatic2002}. Specifically, this requires \mbox{$  \left[2 i \dot{\theta} - (\dot{\phi_{\rm{S}}} - \dot{\phi_{\rm{P}}} + \delta) \sin 2 \theta\right] \rightarrow 0$}
where \(\delta\) is the two-photon detuning, and \(\phi_\mathrm{S},\phi_\mathrm{P}\) represent the phase noise of each laser.
Each term sets a timescale for the evolution of the system, and based on the theory outlined in \cite{yatsenkoDetrimentalConsequencesSmall2014,yatsenkoStimulatedRamanAdiabatic2002} the transfer efficiency $\eta$, becomes dependent on the pulse duration $T$. For balanced pump and Stokes Rabi frequencies $\Omega_P = \Omega_S=\Omega$,
\begin{equation}
    \eta(T) = \exp\left(-\frac{\tau_\mathrm{adi}}{T} -\frac{T}{\tau _{\mathrm{deph}}} \right).\label{Eqn: eta}
\end{equation}
Here, \(\tau_{\mathrm{adi}}\equiv\pi^2\gamma/\Omega^2\) is the timescale of adiabaticity with $\gamma$ being the decay rate from \(\ket{E}\) and \(\tau_{\mathrm{deph}}\) is the dephasing timescale, which includes the effect of laser phase noise and any other noise source that causes detuning between the two lasers. The maximum efficiency achievable is equal to \(\eta(T')=\exp\left(-2\sqrt{\tau_{\mathrm{adi}}/\tau_{\text{deph}}}\right)\) which occurs for a pulse time \(T'=\sqrt{\tau_{\mathrm{adi}} \times \tau_{\mathrm{deph}}}\). The peak efficiency approaches unity as \(\tau_{\mathrm{adi}}\rightarrow 0\) or \(\tau_{\mathrm{deph}}\rightarrow \infty\). 

Decreasing \(\tau_{\mathrm{adi}}\) requires increasing \(\Omega\) by using higher laser intensities. However, this also changes the frequency components of the phase noise which contribute most to $\tau_\mathrm{deph}$. This presents an experimental challenge for STIRAP in molecule formation, where the large (typically, $\sim$\SI{100}{\tera \hertz}) energy gap from Feshbach state to rovibrational ground state is bridged with two independent lasers stabilised to optical cavities or frequency combs. The stabilisation typically produces a servo bump at the \(\sim\!\SI{}{MHz}\) scale which tends to coincide with the STIRAP Rabi frequency. This problem of sensitivity to high-frequency laser phase noise is generic to many platforms where stabilised lasers are used for quantum control, notably trapped ions \cite{nakavEffectFastNoise2023} and Rydberg atoms \cite{grahamRydbergMediatedEntanglementTwoDimensional2019,deleseleucAnalysisImperfectionsCoherent2018}. In this regime, improving state transfer efficiency requires suppression of phase noise at frequencies above the bandwidth of the feedback loop.

We use FF to cancel high-frequency noise with relatively minor modifications to our pre-existing setup. 
Our STIRAP laser system is based on two external cavity diode lasers (ECDLs), each seeding its own fibre amplifier. The Stokes laser operates at \SI{977}{\nm} and the pump at \SI{1557}{\nm}. Both lasers are frequency-stabilised in the standard configuration, using an offset Pound-Drever-Hall (PDH) lock to a high-finesse optical cavity~\cite{dreverLaserPhaseFrequency1983,gregorySimpleVersatileLaser2015}.
To perform FF, we add a time-delay fiber and fiber EOM between each ECDL and their respective fiber amplifier as shown in Fig.\,\ref{fig:concept}(d). We take the error signal from the PDH lock, invert and amplify that signal, and then feed that signal to the EOM. This modulates the light so that any high-frequency phase is cancelled out~\cite{chao_pounddreverhall_2024,liActiveCancellationServoInduced2022}. Effective cancellation relies on matching the amplitude of the modulation with the phase noise on the light. It also requires that the light and the electronic FF signal experience the same time delay between the error signal detection and the modulation. Both of these requirements must be met over the whole bandwidth where noise must be cancelled. 
% The delay fiber is therefore set up to overcompensate the delay associated with the electronics so that the exact time delay can be easily matched by adding coaxial cable \cite{chao_pounddreverhall_2024}. 
A more detailed description of the FF setup is given in the Supplemental Material.

To quantify the performance of FF, we perform self-heterodyne measurements of the phase noise of each laser. We overlap the laser light with light transmitted by the high-finesse cavity and detect the resulting beat note on a photodiode. Here, phase noise at frequencies above the cavity linewidth of \SI{25}{\kHz} can be considered to pertain exclusively to the main laser path~\cite{schmidSimplePhaseNoise2019}. Example measurements using the Stokes laser are shown in Fig.\,\ref{fig:beatnote}. 
In the absence of FF, the servo bump can be seen at $\sim\mkern-4mu\SI{2}{\MHz}$. By adding FF, we significantly reduce the magnitude of phase noise at the servo bump by $\sim\mkern-4mu\SI{20}{\dB}$ such that it becomes comparable to the measurement noise floor. 

When the FF is near optimal it is difficult to measure the true phase noise due to the finite background noise. Therefore to optimise the amplitude and delay of the FF signal we add an artificial phase noise peak at \SI{1}{\MHz} by modulating the current of the laser sinusoidally. The inset of Fig.\,\ref{fig:beatnote} shows the suppression that we measure using this technique as a function of the gain of the inverting amplifier before the EOM (FF gain). We measure a suppression of \SI{29}{\dB} when the amplifier gain is optimal. 

\begin{figure}
    \centering
    \includegraphics[width=\linewidth]{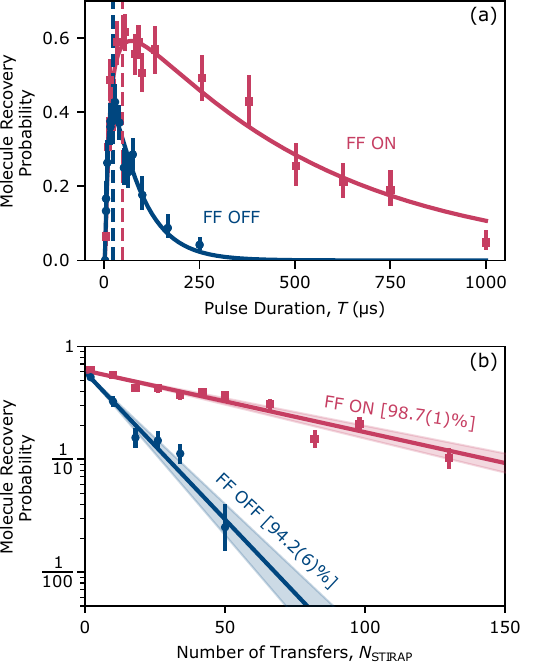}
    \caption{Experimental results showing improved STIRAP using feedforward (FF) (a) Molecule recovery after \NSTIRAP$=10$ with varying STIRAP pulse duration $T$, for FF on (red squares) and FF off (blue circles). Solid lines show fits to the data using the model given in Eq.\,\ref{Eqn: eta}. Dashed vertical lines indicate the optimum pulse durations in each case. (b) Molecule recovery after an even number of STIRAP one-way trips, with and without FF using the optimal pulse durations shown in (a). }
    % The inset illustrates the train of back-to-back STIRAP transfers summing to an even number of $N_{\text{STIRAP}}$.
    \label{fig:results}
\end{figure}
We test the effect of FF on molecular state transfer using single RbCs molecules confined to an array of optical tweezers. The details of this apparatus have been described previously in~\cite{spencePreparation87Rb2022,ruttleyFormationUltracoldMolecules2023,ruttley_enhanced_2024}. 
Rb and Cs atoms in species-specific tweezers are cooled to their respective motional ground states, merged and then associated by ramping over an interspecies Feshbach resonance at~\SI{197.1}{\gauss}~\cite{takekoshiProductionUltracoldGroundstate2012,koppingerProductionOpticallyTrapped2014}, and then transferred to \(\ket{G}\) by STIRAP. We measure the fidelity of STIRAP as part of the overall infidelity in formation of a molecule in \(\ket{G}\) from an atom pair, as explained in \cite{ruttley_enhanced_2024}. If there is an error in forming a molecule and an atom pair remains in the tweezer we detect this via pulling out Rb atoms with a species-specific tweezer into a separate \mbox{\emph{error detection array}}. After the formation of the molecule, we reverse the STIRAP and magneto-association and this time separate the Rb atom into a different \mbox{\emph{imaging array}} of Rb-specific tweezers. Thus we are able to measure the \mbox{\emph{molecule recovery probability}}, \(P_{\mathrm{r}}\) as the probability of imaging Rb and Cs pairs conditioned on not imaging a Rb atom in the error detection array. The STIRAP fidelity is only one of many multiplicative factors in \(P_{\mathrm{r}}\), and so to make measurements more sensitive to STIRAP we perform multiple state transfers on the molecule before dissociation. To improve statistics we prepare arrays, and average over up to six tweezers per shot.

% \jonathan{AG to re-write from here to end of paragraph and add supplemental if required}The molecules are initialised in $\ket{F}$, and we detect the presence of the RbCs molecule in $\ket{F}$ before and after an even number of STIRAP pulses $N_\mathrm{STIRAP}$ to find the \emph{molecule recovery probability} as described in the Supplemental Material. Without STIRAP the molecule recovery probability is limited to $A=60(2)$\% by factors such as the lifetime of the molecules and the fidelity of molecule detection. 

The STIRAP light is focused onto the molecules with waists of \SI{72(3)}{\micro\meter} for the Stokes and \SI{63(3)}{\micro\meter} for the pump. The light propagates perpendicular to the applied magnetic field that defines the quantisation axis. Both lasers are linearly polarised, with the Stokes polarised perpendicular and the pump polarised parallel, to the magnetic field. We can apply up to \SI{110}{\mW} of Stokes light and \SI{272}{\mW} of pump light. By driving one-photon Rabi oscillations as in \cite{molonyProductionUltracold872016} we measure the individual Rabi frequencies of $\Omega^{\mathrm{max}}_\mathrm{S}=1190(30)$\,kHz and $\Omega^{\mathrm{max}}_\mathrm{P}=1170(20)$\,kHz (see Supplemental Material). The intensity of each beam is modulated with acousto-optic modulators to achieve STIRAP using the \(\cos^2\)-pulse shape.

To find the optimal pulse time, we take measurements of the molecule recovery probability after \NSTIRAP$=10$ while varying $T$ as shown in Fig.\,\ref{fig:results} (a). We fit the data with Eq.~\ref{Eqn: eta} to extract values of \Tadi~and \Tdeph. We find good agreement between our results and this model, with a sharp rise in probability below $T'$ as the transfer becomes adiabatic, and then a slow decay from decoherence as $T$ increases further. Comparing FF on and off, we see similar \Tadi ~for both cases as expected.
We find $\Tadi=\SI{1.0(1)}{\micro \second}$ for FF on and $\Tadi=\SI{0.8(1)}{\micro \second}$ for FF off. 
% To compare to Eq. \ref{Eqn: eta}, we assume a linewidth of $\gamma=\SI{35(10)}{\kHz}$~\cite{molonyProductionUltracold872016} from previous measurements, giving a value of \SI{0.25(7)}{\micro \second}. 
In contrast, adding FF changes $\Tdeph$ significantly.  Without FF $\Tdeph=0.73(9)$\,ms, however with FF active this is increased by nearly an order of magnitude to $\Tdeph=5.0(6)$\,ms. This dramatic effect shows that phase noise is indeed the main limiting factor on \Tdeph, and with FF on, the maximum transfer efficiency can peak higher. 

\begin{figure}[t!]
    \centering
    % \def\svgwidth{\columnwidth}
    % \import{Figures/}{ErrorRate_2.pdf_tex}
    \includegraphics[width = \linewidth]{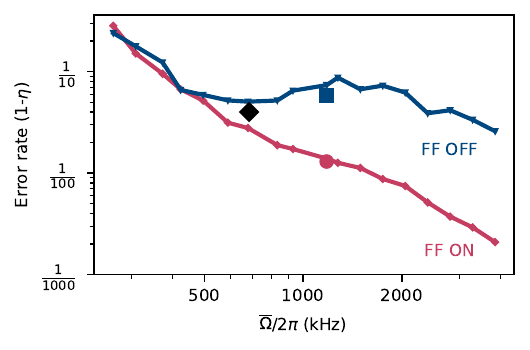}
    \caption{Simulated error rate ($1$-$\eta$) of STIRAP as a function of average maximum Rabi frequency $\overline{\Omega}=(\Omega_{\text{P}}+\Omega_{\text{S}})/2$. A red circle (blue square) marks the experimental data for FF on (off) as measured in Fig.\,\ref{fig:results} (b). The measurement from our previous work in \cite{ruttley_enhanced_2024} is marked by a black diamond. Error bars on the experimental points are smaller than the marker size. }
    \label{fig:error_rate}
\end{figure}

To more precisely determine the STIRAP efficiency, we measure the molecule recovery probability as a function of $N_\mathrm{STIRAP}$ as shown in Fig.\,\ref{fig:results} (b). The transfer time is set to maximise the efficiency, such that without FF $T'=\SI{23.85}{\micro\second}$ and with FF $T'=\SI{45}{\micro\second}$. We perform a log-linear fit to the results with the gradient indicating the efficiency of each passage. The extracted efficiencies are $\eta_{\text{on}}=98.7(1)\%$ for FF on and $\eta_{\text{off}}=94.2(6)\%$ for FF off. These results can be expressed as a reduction in the transfer error by a factor of 4.5(5). As STIRAP efficiency is an important factor in the formation of molecules and is the dominant error in their detection this is an important step in the quantum control of molecules. Bringing the error rate of state transfer below \(\sim\!1\)\% is crucial in many applications including high fidelity quantum simulation \cite{daleyPracticalQuantumAdvantage2022} and quantum information storage\cite{zhangQuantumComputationHybrid2022}.

To understand how higher efficiencies can be achieved, we compare our results to a model of the STIRAP which accounts for the main noise contributions in our system. 
Our simulations are based upon the 3-level model described in~\cite{yatsenkoStimulatedRamanAdiabatic2002}. The laser phase noise was simulated by extracting phase noise in angular units from the cavity-laser beat note spectra in Fig.\,\ref{fig:beatnote} in the same manner as \cite{schmidSimplePhaseNoise2019a}, and then initialising the noise to be the sum of sinusoids matching the spectrum amplitudes and with randomised phases. Lower frequency noise contributions are included in the model as a randomised shot-to-shot two-photon detuning. The largest contribution to this is from magnetic field noise which we estimate causes a deviation of \(\SI{30}{kHz}\). There is also a smaller contribution from the laser linewidth. We measure the linewidth of the Stokes laser to be $\SI{346(3)}{\Hz}$ from the half-width half-maximum of a beat measurement between our laser and an identical but otherwise independent laser system. We assume that the linewidth of the pump and Stokes lasers are the same.  We use the natural decay rate from the excited state which we have previously estimated to be $\gamma=\SI{35(3)}{\kHz}$~\cite{molonyProductionUltracold872016} and neglect coupling to other states. More details of the noise model can be found in the Supplemental Material.

The results of our simulation are presented in Fig.\,\ref{fig:error_rate}, which shows the simulated STIRAP error rate ($1\!-\!\eta$) as a function of the average Rabi frequency $\overline{\Omega}=(\Omega_{\text{P}}+\Omega_{\text{S}})/2$. 
At low Rabi frequencies, the magnetic field noise dominates over the phase noise and efficiency is gained simply by lowering \Tadi, with no advantage to using FF. 
As $\bar{\Omega}$ increases, the transfer becomes more sensitive to laser phase noise. Without FF, this causes the error rate to increase above $\sim\!\SI{700}{\kHz}$, tracing out the shape of the servo bump. With FF on, the error rate decreases as a function of $\bar{\Omega}$, with no apparent impact from the servo bump. 
We plot our measured error rates, along with those from~\cite{ruttley_enhanced_2024} as the markers in Fig.\,\ref{fig:error_rate}. We see good agreement between the model and these experimental measurements. 

Our model indicates that the efficiency can be further improved by increasing $\bar{\Omega}$, which is only limited by the available laser intensity. Extrapolating out to a reasonable value of $\bar{\Omega}=\SI{4}{\MHz}$, we can see the FF has the possibility of opening up an order of magnitude advantage in error rate compared to the FF off and approaches an error rate of 1/1000. This would require an increase in beam intensity of a factor $\sim\!\!20$. This could be easily achieved by tighter focusing of the STIRAP beams to waists sizes of $\SI{35}{\micro\m}$ as used in~\cite{molonyProductionUltracold872016} combined with a modest increase in the laser power to~500\,mW and 300\,mW for the pump and Stokes respectively, which are readily available with increased amplification.

In conclusion, we have demonstrated the use of FF suppression of laser phase noise to significantly enhance state transfer efficiency. For the demanding test case of ground-state transfer using STIRAP in ultracold molecules, we achieve 98.7(1)\% one-way transfer efficiency. This is the highest reported in any ultracold molecule experiment to date. By modelling the transfer in the presence of experimental noise, we find that the efficiency achieved is now limited by the available laser intensity. The simulation implies that efficiencies approaching 99.9\% can be achieved with realistic changes to the beam waist and laser power. Our results enable the state preparation and readout of ultracold molecules with high fidelity which will be crucial for future applications of ultracold polar molecules in quantum simulation and quantum computation. Moreover, we anticipate that the techniques presented here can be readily applied to any optical quantum control scheme that is currently limited by phase noise on a stabilised laser, such as, but not restricted to, qubit operations in trapped ions \cite{akermanUniversalGatesetTrappedion2015a} or Rydberg atoms \cite{deleseleucAnalysisImperfectionsCoherent2018a}.

\bibliographystyle{apsrev4-1} % APS RevTex bibliography style
\bibliography{main}% Produces the bibliography via BibTeX.

\appendix

% \begin{onecolumngrid}

\section{Supplemental Material}

\subsection{Laser Setup and Feedforward Noise Suppression}

\begin{figure}
    \centering
    \includegraphics[width = \columnwidth]{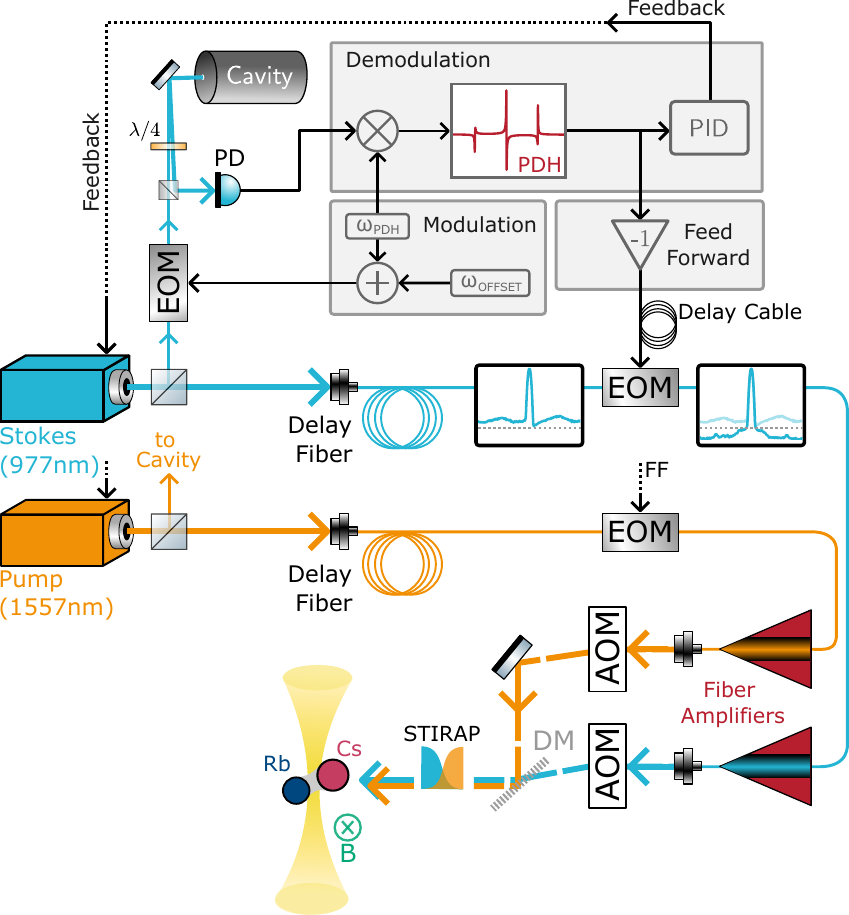}
    \caption{Detailed diagram of the experimental setup. 
    Lasers are locked to the cavity via the PDH technique and the high-frequency phase noise is fed forward onto the FF EOM. A delay fibre is used to create an optical delay to match the timing of the FF signal delivery. Each beam is then fibre amplified and passed through an AOM before being spatially overlapped with each other on a dichroic mirror (DM). The direction of the bias magnetic field (B) is orthogonal to the propagation direction of the STIRAP beams as shown.} 
    \label{fig:sup laser setup}
\end{figure}

The laser system is illustrated in Fig.\,\ref{fig:sup laser setup}. Excluding FF the setup is identical to that presented in \cite{gregorySimpleVersatileLaser2015} and uses an offset PDH lock to allow tuning to the molecular transitions. A high finesse ($\sim$\!\! 30,000) optical cavity (Stable Laser Systems) coated for both pump and Stokes wavelengths is used, rendering any vibrational noise on the cavity common to both lasers. For each input to the cavity, a waveguide electro-optic modulator (EOM) introduces the PDH modulation, $\omega_{\text{PDH}}\:(\sim\!\SI{20}{\MHz})$ and an offset modulation $\omega_{\text{offset}}\:(\sim\!\SI{100}{\MHz})$  for tuning to the STIRAP transitions \cite{gregorySimpleVersatileLaser2015}. Locking to one of the offset sidebands detunes the laser by $\omega_{\text{offset}}$ from the cavity mode. The cavity reflection is focused onto an InGaAs fast photodiode (\SI{150}{\MHz}), and the signal is then demodulated at $\omega_{\text{PDH}}$ to produce an error signal that is split for feedback via a PID controller.

Since the cavity will reflect any frequency components outside of the linewidth, the reflection used for the PDH error signal contains the phase noise information at high frequency, forming the basis of the FF signal as explained in \cite{chao_pounddreverhall_2024}. We split off the PDH error signal after demodulation and use a voltage-controlled amplifier in inverting configuration (VCA810 Texas Instruments, \SI{3}{\decibel} bandwidth of \SI{12}{\MHz}) to drive a waveguide EOM ( iXblue NIR-MPX-LN-0.1 for \SI{997}{\nm}, iXblue MPX-LN-0.1 for \SI{1557}{\nm}). We opted to use fiber EOMs to simplify the drive electronics as the required \(V_\pi\) is small, however, the FF \begin{figure}[t!]
    \centering
    \includegraphics[width =\columnwidth]{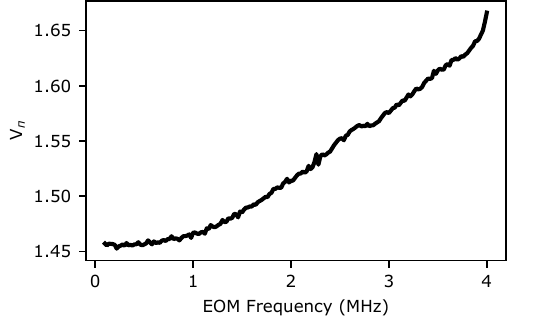}
    \caption{Response of iXblue NIR-MPX-LN-0.1 EOM measured by observing a phase modulation sideband on the beat note spectrum after applying pure-tone phase modulation.   }
    \label{fig:eom}
\end{figure}
technique only requires peak-to-peak phase modulation on the scale of the phase noise of the stabilised laser so we expect that similar performance may be achievable with free space EOMs. For feedforward to work over a wide bandwidth the the EOM \(V_\pi\) must be as flat as possible \cite{chao_pounddreverhall_2024,liActiveCancellationServoInduced2022, heungjaechoiEfficiencyEnhancementFeedforward2010}.  These EOMs are designed for near DC operation. We measure the response by observing optical sidebands, results are shown in Fig.\,\ref{fig:eom}. The ideal system would use an amplifier with a tuned gain profile to flatten the frequency response of the EOM, however we found this was not necessary to see significant noise cancellation. 

Delay matching between the PDH FF error signal and the optical path is critical to effective FF. We first estimated the delay in the PDH setup by setting up the FF with optical delay fiber and additional electronic delay until the FF signal phase delay is approximate \(2\pi\) and FF is able to weakly cancel the phase noise \cite{liActiveCancellationServoInduced2022}. This gave an estimated electronic delay of \SI{55(3)}{\nano \second}. We then selected an optical delay length greater than the estimated delay in the PDH arm, such that the difference between the paths could be optimised by adding electronic delay to the FF signal with coaxial cable.

After passing through the FF fibre EOM, the optical power is boosted by fibre amplifiers from Precilasers. The fibre amplifiers then output to free-space and AOMs are utilised to perform the intensity ramping necessary for the pulse shape. A dichroic mirror is used to spatially overlap the pump and Stokes beams. Then a final mirror directs the light through an achromatic lens, focusing the beams down onto the tweezer array. We measured the phase noise spectrum of the amplified light using the cavity beat note technique and saw no significant increase in noise after amplification.

% \subsection{Preparation of RbCs in Tweezer arrays and measurement of Molecule recovery probability}

% Two spatially-offset, state-selective tweezer arrays trap single Rb and Cs molecules at wavelengths of \SI{817}{\nm} and \SI{1066}{\nm} respectively. Raman sideband cooling then prepares the atoms in the respective motional groundstates, $(F\!=\!1,m_F\!=\!1)_\text{Rb}$ and $(F\!=\!3,m_F\!=\!3)_\text{Cs}$. The tweezers are then merged and the \SI{817}{\nm} tweezer is ramped off, leaving \SI{1066}{\nm} tweezers with Rb+Cs pairs. A bias magnetic field is then adiabatically swept across an inter-species Feshbach resonance at \SI{197.1}{\gauss}\cite{koppingerProductionOpticallyTrapped2014}

% \jonathan{Explain in detail molecule recovery probability and effect of different lifetimes on the many pulse measurements}

\subsection{STIRAP pulses}
\begin{figure}[t!]
    \centering
    \includegraphics[width =\columnwidth]{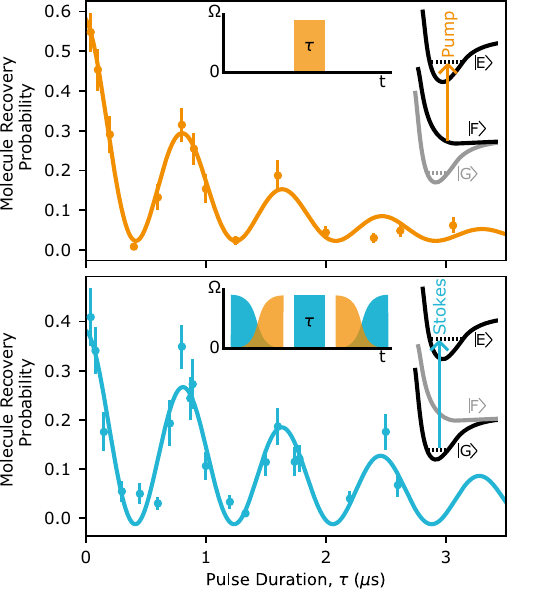}
    \caption{Rabi flopping on the pump $\ket{F} \rightarrow \ket{E}$ and Stokes $\ket{G} \rightarrow \ket{E}$ transitions after a square pulse excitation of duration $\tau$. Solid lines show fits to a decaying oscillation model used to extract the Rabi frequency. Inset: one-photon transitions being coupled, along with the pulse sequence used to produce the flop.  }
    \label{fig:RabiFlop}
\end{figure}

 For a one-way STIRAP transfer from $\ket{F} \rightarrow \ket{G}$ the pulse shapes for pump and Stokes are 
\begin{equation} \label{eq1}
\begin{split}
\Omega_{\rm{P}}(t) & = \Omega_{\rm{P}} \left[1-\cos^2{\left( \frac{\pi}{2} \frac{t}{T} \right)}\right], \\
\Omega_{\rm{S}}(t) & = \Omega_{\rm{S}} \cos^2{\left(\frac{\pi}{2}\frac{t}{T} \right)},
\end{split}
\end{equation}
where $T$ defines the pulse duration. To measure $\Omega_{\rm{P}}$ and $\Omega_{\rm{S}}$, we drive one-photon transitions with a square pulse excitation on the pump and Stokes transitions to produce a Rabi flop as seen in Fig.\,\ref{fig:RabiFlop}. For the pump transition $\ket{F} \rightarrow \ket{E}$, this is straightforward as the molecules should cycle back to $\ket{F}$ where they can be detected. However for the Stokes transition $\ket{G} \rightarrow \ket{E}$, a STIRAP transfer is required to initialise them in $\ket{G}$, then after the square pulse, a reverse STIRAP transfer back to $\ket{F}$ for detection. The Rabi flop is then fitted to a decaying oscillation model to determine the Rabi frequency.

\subsection{Modelling of STIRAP}

Following \cite{yatsenkoStimulatedRamanAdiabatic2002} we model the STIRAP process using the Hamiltonian
\begin{widetext}
\begin{equation}
H_{\text{BD}} = \frac{h}{2}
\begin{pmatrix}
2 \dot{\phi_{\rm{S}}} \sin ^{2} \theta + 2 \dot{\phi_{\rm{P}}} \cos ^{2} \theta + \delta \cos 2 \theta & \Omega_{\rm{rms}} & 2 i \dot{\theta} - (\dot{\phi_{\rm{S}}} - \dot{\phi_{\rm{P}}} + \delta) \sin 2 \theta \\
\Omega_{\text{rms}} & -2 \Delta - i \gamma & 0 \\
-2 i \dot{\theta} - (\dot{\phi_{\rm{S}}} - \dot{\phi_{\rm{P}}} + \delta) \sin 2 \theta & 0 & 2 \dot{\phi_{\rm{S}}} \sin ^{2} \theta + 2 \dot{\phi_{\rm{P}}} \cos ^{2} \theta - \delta \cos 2 \theta
\end{pmatrix},
\label{eq:HBD}
\end{equation}
\end{widetext}
expressed in the bright-excited-dark state basis $\{B,E,D\}$ with time dependence notation suppressed. Here $\Delta$ and $\delta$ represent the one- and two-photon detunings respectively and $\Omega_{\text{rms}} = \sqrt{\Omega_{\rm{P}}^2 + \Omega_{\rm{S}}^2}$. We numerically propagate this Hamiltonian using the QuTiP package, inputting the real phase noise for both lasers and a realistic model for the magnetic field instability inferred from in-situ measurement. The phase noise for each laser is measured via self-heterodyne measurements with the cavity transmission, as in Fig.\,\ref{fig:beatnote}, with only frequencies above the cavity linewidth considered in the model. The phase noise is then added to the model by initialising pure-tones at each frequency interval in the measured spectrum with its corresponding amplitude and a randomised initial phase.  The magnetic-field noise at \SI{2}{\kHz} was measured to be \SI{4}{\milli \gauss} 
by measuring the broadening of a microwave transition from $\ket{F=1,m_F=1}$ to $\ket{F=2,m_F=2}$ with Rb atoms in the tweezer array under the same conditions as the STIRAP measurements. The spectrum of the magnetic noise was estimated to have a $1/f$ character and added to the model as a one- and two-photon detuning in the same manner as the laser phase noise.  The linewidth of the lasers was inferred from a beat note of the Stokes laser with another independent but identically configured Stokes laser system, giving $\SI{346(3)}{\Hz}$. The contribution of the linewidth is added as a randomised shot-to-shot one- and two-photon detuning, which is normally distributed about zero with a standard deviation of \SI{346}{\Hz}. For each simulated Rabi frequency, the ratio of $\Omega_{\rm{P}}$ and $\Omega_{\rm{S}}$ is held consistent with the experiment and multiplied by a factor. The pulse duration is varied for each Rabi frequency and with the maximum efficiency extracted numerically.
% \end{onecolumngrid}

% For the Stokes flop, a STIRAP transfer initialises the molecule in $\ket{G}$ then the square pulse is applied for the flop after whihc a reverse STIRAP transfer puts the molcule back into $\ket{F}$ for detection. 
\end{document}

\section{STIRAP Model}

\section{PDH Lock}

A high finesse ($\sim$30,000) optical cavity is used for servoing of the laser frequency via PDH. Since the cavity will reflect any frequency components outside of the linewidth, the reflection used for the PDH error signal contains the phase noise information at high frequency, forming the basis of the FF signal \cite{chaoPoundDreverHallFeedforwardLaser2023}. 
Optics in the cavity are coated for both pump and stokes wavelengths, facilitating dichroic operation and rendering any cavity noise common to both lasers. For each input to the cavity, an electro-optic modulator (EOM) introduces the PDH modulation, $\omega_{\text{PDH}}\:(\sim \SI{20}{\MHz})$ and an offset modulation $\omega_{\text{offset}}\:(\sim \SI{100}{\MHz})$  for tuning to the STIRAP transition \cite{gregorySimpleVersatileLaser2015}. Locking to one of the $\omega_{\text{offset}}$ then effectively detunes the laser by $\omega_{\text{offset}}$. The cavity reflection is focused onto an InGaAs fast photodiode (\SI{150}{\MHz}), the signal is then demodulated at $\omega_{\text{PDH}}$ to produce an error signal that is split for feedback via a PID controller,

\section{Preperation of RbCs}
Briefly,two spatially-offset, state-selective tweezer arrays trap single Rb and Cs molecules at wavelengths of \SI{817}{\nm} and \SI{1066}{\nm} respectively. Raman sideband cooling then prepares the atoms in the respective motional groundstates, $(F\!=\!1,m_F\!=\!1)_\text{Rb}$ and $(F\!=\!3,m_F\!=\!3)_\text{Cs}$. The tweezers are then merged  and the \SI{817}{\nm} tweezer is ramped off, leaving \SI{1066}{\nm} tweezers with Rb+Cs pairs. A bias magnetic field is then adiabatically swept across an inter-species Feshbach resonance at \SI{197.1}{\gauss}\cite{koppingerProductionOpticallyTrapped2014}